\documentclass[%
 reprint,
 amsmath,amssymb,
 aps,
]{revtex4-1}

\usepackage{graphicx}
\usepackage{color}
\usepackage{dcolumn}
\usepackage{bm}
\usepackage{lineno,hyperref}
\usepackage{epsfig}
\usepackage{float}
\usepackage{bookmark}
\usepackage{times}
\usepackage{amsmath}
\usepackage{color}
\usepackage{epstopdf}

\begin{document}

\preprint{APS/123-QED}

\title{Evolutionary dynamics in the public goods games with switching between punishment and exclusion}

\author{Linjie Liu}

\author{Shengxian Wang}

\author{Xiaojie Chen}
\email{xiaojiechen@uestc.edu.cn}
\affiliation{School of Mathematical Sciences, University of Electronic Science and Technology of China, Chengdu 611731, China}

\author{Matja{\v z} Perc}
\email{matjaz.perc@uni-mb.si}
\affiliation{Faculty of Natural Sciences and Mathematics, University of Maribor, Koro{\v s}ka cesta 160, SI-2000 Maribor, Slovenia}
\affiliation{Complexity Science Hub, Josefst{\"a}dterstra{\ss}e 39, A-1080 Vienna, Austria}
\affiliation{School of Electronic and Information Engineering, Beihang University, Beijing 100191, P.R. China}

\begin{abstract}
Pro-social punishment and exclusion are common means to elevate the level of cooperation among unrelated individuals. Indeed, it is worth pointing out that the combined use of these two strategies is quite common across human societies. However, it is still not known how a combined strategy where punishment and exclusion are switched can promote cooperation from the theoretical perspective. In this paper, we thus propose two different switching strategies, namely peer switching that is based on peer punishment and peer exclusion, and pool switching that is based on pool punishment and pool exclusion. Individuals adopting the switching strategy will punish defectors when their numbers are below a threshold and exclude them otherwise. We study how the two switching strategies influence the evolutionary dynamics in the public goods game. We show that an intermediate value of the threshold leads to a stable coexistence of cooperators, defectors and players adopting the switching strategy in a well-mixed population, and this regardless of whether the pool-based or the peer-based switching strategy is introduced. Moreover, we show that the pure exclusion strategy alone is able to evoke a limit cycle attractor in the evolutionary dynamics, such that cooperation can coexist with other strategies.
\end{abstract}

\maketitle

\textbf{Large-scale cooperation among unrelated individuals distinguishes humans markedly from other animals, and it is indeed crucial for our evolutionary success. Cooperation is remarkable because it is costly for the individual that cooperates, but it is beneficial for the society as a whole. As such, to cooperate is in contradiction with the Darwinistic principle of maximizing personal fitness, and it is therefore challenged by defection. Individuals are thus torn between what is best for them and what is best for the society -- the blueprint of a social dilemma. The theoretical framework for studying this fascinating aspect of our biology is evolutionary game theory, with the most commonly used games being the prisoner's dilemma and the public goods game. The latter two games describe the essence of the problem succinctly for pairwise and group interactions, respectively. The resolution of social dilemmas towards the pro-social outcome has received ample attention in the recent past. In our paper, we extend the scope of the classic public goods game with cooperators and defectors to account for a new third type of strategy, namely the switching strategy that either punishes or excludes defectors depending on their numbers. Our research reveals fascinatingly different evolutionary dynamics, including the stable coexistence of three strategies and limit cycles, which enables cooperators to survive where otherwise they would perish. These results have important implications for the better understanding of cooperation, and we also hope they will be inspiring for economic experiments in the future.}

\section{Introduction}
The emergence of altruistic behavior among unrelated individuals has been a puzzling phenomenon, since such behavior is usually costly to perform but benefits others
\cite{AxelrodR81,Dur05,Tanimoto07,Tanimoto05,Santos08,Wang11,nowakm04,zhang18,chenxj08,perc08,perc10,perc17}.
A number of game theoretic researches over the past decades have provided numerous answers to this question, such as indirect reciprocity, reputation, reward, and punishment \cite{nowakm06,fu08,ded04,chenx13,szolnoki10,szolnoki12,wang13,chenx15}. Among them, scholars pay more attention to the role of pro-social punishment played in promoting public cooperation \cite{Szolnoki11,perc_12_njp,chenx14,chenxj14}. Theoretical and experimental studies have indicated
that pro-social punishment can reduce the expected payoff of self-interested individuals since they need to pay an extra fine \cite{rand09,gurerkq06,fehre02}. However, such action is also costly, it results in the second-order free-riders dilemma, in which individuals may prefer to benefit from punishment but do not contribute the related costs
\cite{yang2018,wangz13,szolnoki17,perc15,chenxj15,bodyr10,sasaki15,Szolnoki16_njp}.

Social exclusion has recently drawn more attention since it can perform better than costly punishment for maintaining cooperation  \cite{sasaki13,lik15,szolnokia17,lik16,liul17,sui18}.
It is thought that social exclusion is still an advantageous strategy even facing with a large number of free-riders,
due to the fact that excluding defectors from sharing benefit can decrease the number of beneficiaries.
Recently, such exclusion strategy has been studied from an evolutionary perspective by Sasaki and Uchida \cite{sasaki13}. The results show that social exclusion strategy can overcome two difficulties of costly punishment: first, rare punishers can not defeat a large area of free-riders; second, punishers will be eliminated by natural selection in the presence of second-order free-riders. Subsequently, Liu et al. \cite{liul17} studied the competition between pro-social exclusion and punishment in finite populations, and claimed that social exclusion can always do better than punishment.

It is worth pointing out that in most previous studies the evolution of these two incentive strategies has been explored in a manner in which the two strategies work independently \cite{sasaki13,liul17}. But in the realistic world, what is pretty widespread is the combined use of these sanctioning strategies. It is still unclear how such combined strategy can promote cooperation from the theoretical perspective. Considering the different roles of pro-social punishment and exclusion played in raising cooperation, it is interesting to investigate how to jointly use the two strategies for the promotion of cooperation. Notice that a specific example in the realistic world is the problem of environmental pollution control. If the number of enterprises discharging pollutants illegally exceeds a given threshold, the Environmental Protection Agency (EPA) will force these enterprises to suspend operations. Otherwise the EPA will impose fines on them \cite{hellande98,shimshackj05}.

In this paper, we thereby propose a switching strategy with which individuals can either punish or exclude free-rider in the public goods game (PGG) depending on the number of defectors in the group. Specifically, if the number of defectors in the group is above a certain threshold, individuals who contribute to the monitoring organization will behave as excluders, otherwise they act as punishers. We consider these assumptions and then construct a model based on the one in our previous work \cite{liulj17}. Furthermore, in our model we further consider that strategy-switching individuals need to pay a monitoring cost, which is ignored in Ref. \cite{liulj17}, but it is more reasonable for real-world systems. This is because that monitoring the whole population and knowing the number of free-riders in the population require a certain monitoring cost. We find that a middle threshold can make the system converge to a stable coexistence state of cooperators, defectors, and strategy switching players no matter whether pool-based switching strategy or peer-based switching strategy is used. In addition, we prove that when pure exclusion strategy is considered into public goods games, the population system can exhibit a limit cycle where cooperative strategy can coexist with other types of strategies.

\section{Model and Method}\label{section2}

\subsection{Public goods game}

We consider an infinite well-mixed of individuals who play the public
goods game. In a group of $N$ individuals, each player has the
opportunity to cooperate by contributing to the common pool with a cost $c$ to itself, or act as defectors by contributing nothing.
Then the sum of all contributions in each group is
multiplied by an enhancement factor $r$ ($1<r<N$) and equally
distributed among all the $N$ individuals. Thus if all individuals choose to cooperate, the group can yield the maximum benefit $rc-c$ for each player.
If all choose to defect, the group can get nothing.

\subsection{Switching strategy of pro-social punishment and exclusion}

We introduce a switching strategy to depict the combined use of pro-social punishment and exclusion.
Different forms of punishment strategies can be chosen based on whether the number of defectors in the group exceeds the tolerance threshold.
Concretely, if the number of free-riders in the group exceeds a given threshold $T$, those players with the switching strategy will exclude free-riders, otherwise punish them.
Thus, the levels of tolerance threshold are determined by the number of defectors in the group, namely, $T=0, \cdots, N$.
In particular, $T=0$ means that all free-riders will be excluded from sharing public goods once there exist strategy-switching players,
while $T=N$ means that all defectors will be punished.
Here, we introduce two switching strategy, namely, peer-based switching and pool-based switching. It is necessary to point out that regardless of peer-based switching strategy or pool-based switching strategy, strategy switching individuals need to bear a monitoring cost $\tau$ for detecting the number of defectors in the group before they exclude or punish free-riders.

\subsubsection{Peer-based switching strategy}

Here we consider peer-based switching strategists who not only contribute to the common pool but also monitor the number of free-riders in the group.
If the level of defection exceeds a certain level in the group, they become excluders who prevent defectors collecting benefit from the
public goods sharing at a cost $c_{E}$ on every defector in the group, otherwise they change to be punishers who impose a fine $\beta$ on each free-rider at a cost $\gamma$ for themselves \cite{sigmundk10}.
Then the payoffs of pure cooperators ($C$), pure defectors ($D$), and peer-based switching strategists ($I_{E}$) obtained from the group can be respectively given by
\setlength{\arraycolsep}{0.0em}
\begin{eqnarray}
\pi_{C}&=&\left\{
\begin{aligned}
&rc-c,  \quad  \text{if}\ N_{I_{E}}\neq0\ \text{and} \ N_{D}\geq T;\\
&\frac{rc(N_{C}+N_{I_{E}}+1)}{N}-c,     \quad   \text{otherwise}
\end{aligned}
\right.\\
\pi_{D}&=&\left\{
\begin{aligned}
&0,  \quad  \text{if}\ N_{I_{E}}\neq0\ \text{and} \ N_{D}\geq T;\\
&\frac{rc(N_{C}+N_{I_{E}})}{N}-N_{I_{E}}\beta ,    \quad    \text{otherwise}
\end{aligned}
\right.\\
\pi_{I_{E}}&=&\left\{
\begin{aligned}
&rc-c-N_{D}c_{E}-\tau,  \quad  \text{if} \ N_{D}\geq T;\\
&\frac{rc(N_{C}+N_{I_{E}}+1)}{N}-c-N_{D}\gamma-\tau ,   \quad  \text{otherwise}
\end{aligned}
\right.
\end{eqnarray}
\setlength{\arraycolsep}{5.0pt}
where $N_{C}, N_{D},$ and $N_{I_{E}}$ denote the numbers of cooperators, defectors, and peer-based switching strategists among the other $N-1$ players, respectively.

\subsubsection{Pool-based switching strategy}

Different from peer-based switching strategy, pool-based switching strategists resort to the institution
of monitoring which can choose to exclude defectors or punish them by giving a corresponding fine $B$. The costs of exclusion and punishment are $\delta$ and $G$, respectively. Accordingly, the payoffs of cooperators, defectors, and pool-based switching strategists ($I_{F}$) obtained from the group can be respectively written as follows.
\setlength{\arraycolsep}{0.0em}
\begin{eqnarray}
\pi_{C}&=&\left\{
\begin{aligned}
&rc-c,  \quad  \text{if}\ N_{I_{F}}\neq0\ \text{and} \ N_{D}\geq T;\\
&\frac{rc(N_{C}+N_{I_{F}}+1)}{N}-c,     \quad   \text{otherwise}
\end{aligned}\label{function1}
\right.\\
\pi_{D}&=&\left\{
\begin{aligned}
&0,  \quad  \text{if}\ N_{I_{F}}\neq0\ \text{and} \ N_{D}\geq T;\\
&\frac{rc(N_{C}+N_{I_{F}})}{N}-N_{I_{F}}B ,    \quad    \text{otherwise}
\end{aligned}\label{function2}
\right.\\
\pi_{I_{F}}&=&\left\{
\begin{aligned}
&rc-c-\delta-\tau,  \quad  \text{if} \ N_{D}\geq T;\\
&\frac{rc(N_{C}+N_{I_{F}}+1)}{N}-c-G-\tau ,     \quad   \text{otherwise}
\end{aligned}
\right.
\end{eqnarray}
\setlength{\arraycolsep}{5.0pt}
where $N_{C}, N_{D},$ and $N_{I_{F}}$ denote the numbers of cooperators, defectors, and pool-based switching strategists among the $N-1$ players, respectively.

\subsection{Replicator dynamics}

In a well-mixed population the fraction of $C, D$, and $I (I_{E}$ or $I_{F}$) players can be denoted by $x, y,$ and $z$, respectively.
Thus, $x, y, z\geq0$ and $x+y+z=1$. Consequently,
the strategy evolution can be studied by using replicator
equations \cite{schuster83,hofbauer03,Hauert2002,Sasaki2011,Tanimoto15,wang2018,Rand11}
\begin{eqnarray}
\left\{
\begin{aligned}
\dot{x}&=x(P_{C}-\bar{P}),  \\
\dot{y}&=y(P_{D}-\bar{P}),  \\
\dot{z}&=z(P_{I}-\bar{P}),
\end{aligned}
\right.
\end{eqnarray}
where $P_{C}, P_{D}$, and $P_{I}$ represent the expected payoffs of $C, D,$ and $I$, respectively, and $\bar{P}=xP_{C}+yP_{D}+zP_{I}$ represents the average payoff of the whole population. And we have
\begin{eqnarray}
P_{i}&=&\sum_{N_{C}=0}^{N-1}\sum_{N_{D}=0}^{N-N_{C}-1}\binom{N-1}{N_{C}}\binom{N-N_{C}-1}{N_{D}}\nonumber\\
&&x^{N_{C}}y^{N_{D}}z^{N-N_{C}-N_{D}-1}\pi_{i},
\end{eqnarray}
where $i=C$, $D$, or $I$.

To better characterize the evolutionary dynamics of the population for different switching threshold, we present our results regarding the switching threshold in the form of bifurcation diagrams in the following section. In addition, we provide detailed theoretical analysis when the switching threshold is set to $N$ or $0$. Unless otherwise specified, theoretical analyses in special conditions are presented in Appendix A, B, C, and D, respectively.

\section{Results}
\subsection{Evolutionary dynamics in the population with peer-based switching strategy}
\begin{figure*}[!t]
\begin{center}
\includegraphics[width=15cm]{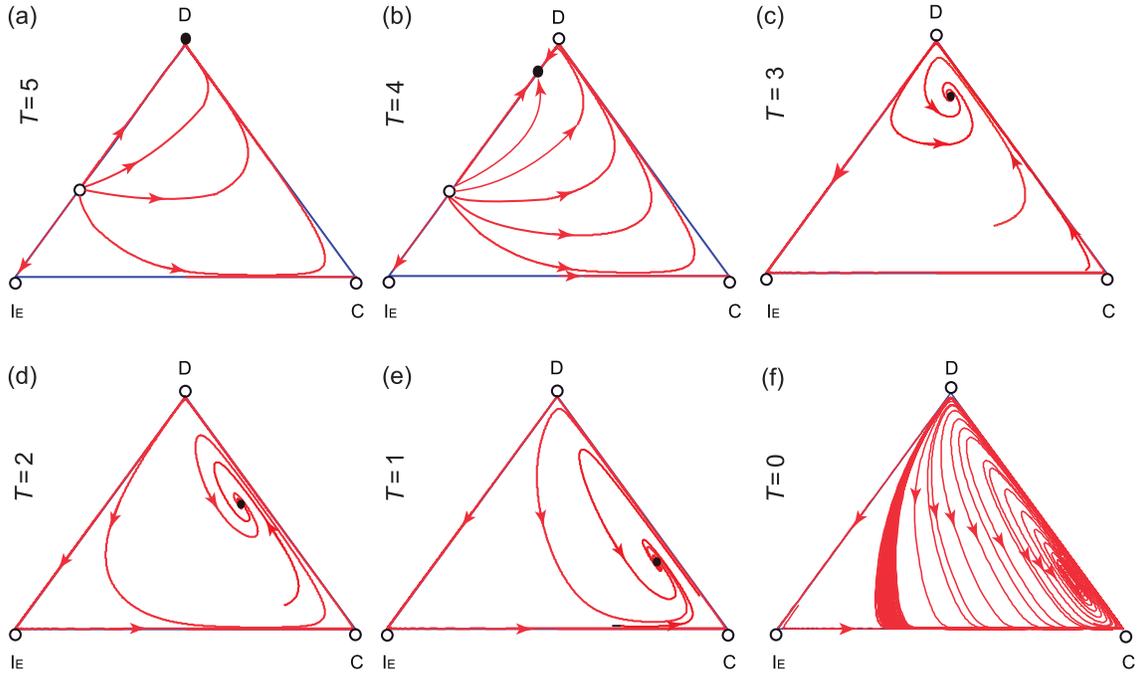}
\caption{Effects of peer-based switching strategy on cooperation for different switching threshold $T$.
The triangles represent the state space $S_{3} =\{(x, y, z): x, y, z\geq0,$ and $x+y+z=1\}$, where $x, y,$ and $z$ are the frequencies of $C, D,$ and $I_{E}$, respectively. Filled circles
represent stable fixed points whereas open circles represent unstable
fixed points. The threshold values are (a) $T=5$, (b) $T=4$, (c) $T=3$, (d) $T=2$, (e) $T=1$, and (f) $T=0$, respectively. Other parameters: $N=5$, $r=3$, $c=1$, $\beta=\gamma=0.4$, $\tau=0.1$, and $c_{E}=0.4$. }\label{fig1}
\end{center}
\end{figure*}
We first present the results of evolutionary dynamics in the population with peer-based switching strategy for
different values of $T$, as shown in Fig.\ref{fig1}. Clearly, if level of tolerance is strong enough ($T=N$), players will opt for defection
(more detailed theoretical analysis is shown in the Appendix A.1).
Besides, an unstable equilibrium point can appear on edge $I_{E}D$ (see Fig.\ref{fig1}(a) and (b)). It needs to stress that a specific stable point appears on the edge $I_{E}D$ when we decrease threshold value slightly (see Fig.\ref{fig1}(b)).
For an intermediate threshold ($T=3$), a new dynamical characteristic appears, that is, an interior stable equilibrium point displays the simplex $S_{3}$.
With decreasing $T$, the interior stable point moves along a straight line approximately parallel to the edge CD (see Fig.\ref{fig1} (c), (d), and (e)).
If $T$ decreases still further to $T=0$, peer-based switching strategist will always use the exclusion action,
Interestingly, defectors will be excluded by peer excluders, defectors dominate cooperators, and cooperators invade peer excluders, forming a heteroclinic cycle on the boundary of simplex $S_3$. In order to judge the stability of this heteroclinic cycle, we calculate the eigenvalues of the Jacobian matrix of the three vertex equilibrium points as follows,
\begin{eqnarray}
\left\{
\begin{aligned}
\lambda_{I_{E}}^{-}&=-(rc-c-\tau),\quad \lambda_{I_{E}}^{+}=\tau, \\
\lambda_{C}^{-}&=-\tau,\quad \lambda_{C}^{+}=c-\frac{rc}{N}, \\
\lambda_{D}^{-}&=\frac{rc}{N}-c,\quad \lambda_{D}^{+}=rc-c-\tau-(N-1)c_{E}.
\end{aligned}
\right.
\end{eqnarray}
Then we define that $\lambda_{I_{E}}=-\frac{\lambda_{I_{E}}^{-}}{\lambda_{I_{E}}^{+}}, \lambda_{C}=-\frac{\lambda_{C}^{-}}{\lambda_{C}^{+}},$ and $\lambda_{D}=-\frac{\lambda_{D}^{-}}{\lambda_{D}^{+}}$, and we have $\lambda=\lambda_{I_{E}}\lambda_{C}\lambda_{D}=\frac{rc-c-\tau}{rc-c-\tau-(N-1)c_{E}}>1$. Thus the heteroclinic cycle is asymptotically stable \cite{park18} (see Appendix B.3 for theoretical analysis).
Besides, a stable limit cycle exists in the interior of the simplex $S_{3}$ (see Appendix B.4 for theoretical analysis). Frequencies of three strategies oscillate, but the interior equilibrium point is unstable (see Fig.\ref{fig1}(f), and more detailed theoretical analysis is presented in the Appendix B.2).
\begin{figure*}[!t]
\begin{center}
\includegraphics[width=15cm]{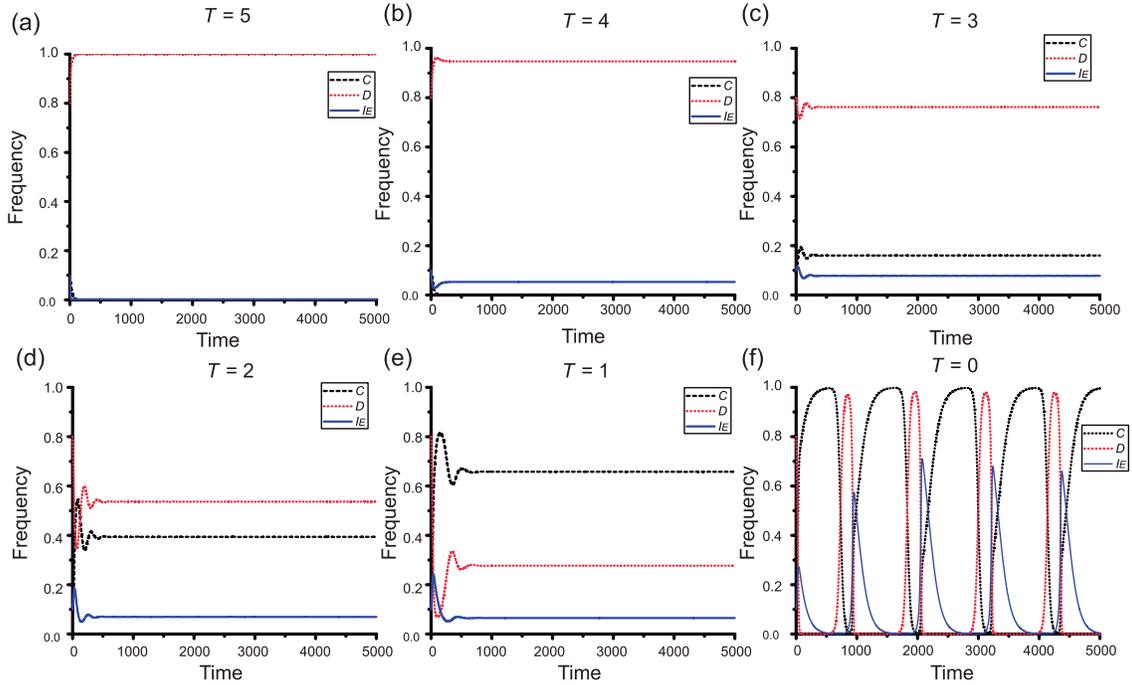}
\caption{Time evolution of the fractions of three strategies $C$ (black dash line), $D$ (red dot line), and $I_{E}$ (blue solid line) for different switching threshold values $T$. Initial conditions: $(x, y, z)=(0.1, 0.8, 0.1)$. Parameters: $N=5$, $r=3$,
$c=1$, $\beta=\gamma=0.4$, $\tau=0.1$, and $c_{E}=0.4$. The threshold values are (a) $T=5$, (b) $T=4$, (c) $T=3$, (d) $T=2$, (e) $T=1$, and (f) $T=0$, respectively. }\label{fig2}
\end{center}
\end{figure*}

In order to shed light on the details behind the results presented in Fig.\ref{fig1},
we depict the frequency of the mentioned three strategies as a function of time for different switching thresholds in Fig.\ref{fig2}. It can be observed that although pure defectors can always occupy the highest proportion of the population for $T\geq 2$, the evolutionary advantage of defectors is weakened gradually with decreasing $T$ (see Fig.\ref{fig2}(a-d)). However, it is worth noting that when $T=1$, pure cooperators have higher fitness than defectors (see Fig.\ref{fig2}(e)). In particular, a periodic oscillation occurs when peer exclusion is performed (see Fig.\ref{fig2}(f)), which is corresponding to the stable limit cycle shown in Fig.1(f).

\subsection{Evolutionary dynamics in the population with pool-based switching strategy}

\begin{figure*}[!t]
\begin{center}
\includegraphics[width=15cm]{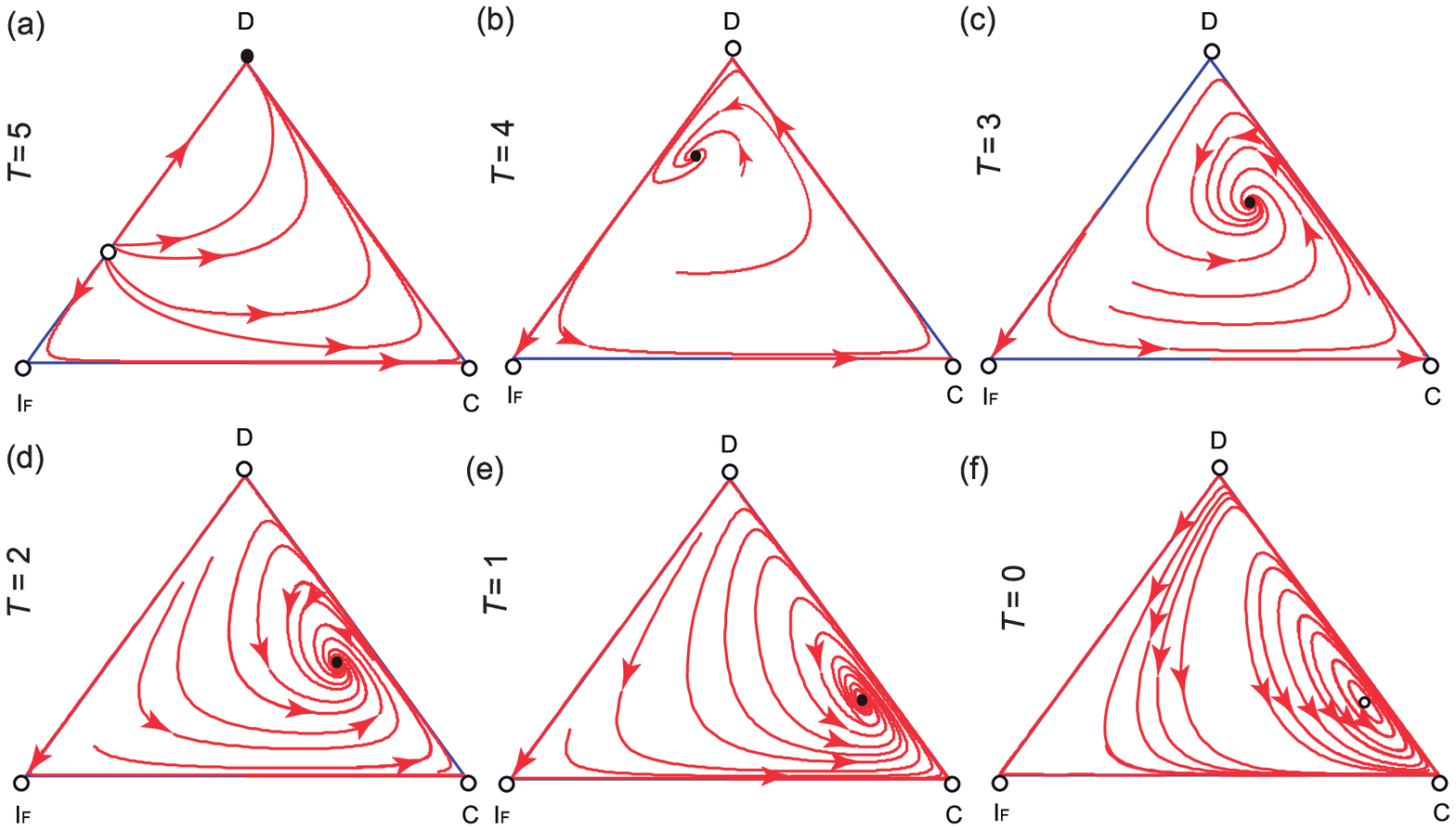}
\caption{Effects of pool-based switching strategy on cooperation for different switching threshold $T$. Parameters: $N=5$, $r=3$,
$c=1$, $B=0.4$, $G=\delta=0.4$, and $\tau=0.1$. The threshold values are (a) $T=5$, (b) $T=4$, (c) $T=3$, (d) $T=2$, (e) $T=1$, and (f) $T=0$, respectively.}\label{fig3}
\end{center}
\end{figure*}
We next illustrate how the introduction of pool-based switching strategy influences the cooperation level for different $T$, as shown in Fig.\ref{fig3}.
When $T=N$, pool-based switching strategists will always act as pool punishers. In this case, no interior equilibrium point appears in the simplex $S_{3}$ since $P_{I_{F}}<P_{C}$, while an unstable equilibrium point exists on edge $I_{F}D$ with $z=\frac{N(c+\delta+\tau)-rc}{N(N-1)B}$ (for further details, see Appendix C.1). The system only has one stable point $D$, which means that defectors will dominate the whole population.
However, this phase portraits will be changed when we reduce the tolerance threshold $T$. When $T=4$, an interior stable equilibrium point is present in the simplex $S_{3}$, thus cooperators, defectors, and pool-based switching strategists can coexist steadily in the population (see Fig.\ref{fig3}(b)). Furthermore, with decreasing $T$ the interior stable equilibrium point moves towards full cooperation state (see Fig.\ref{fig3}(b), (c), (d), and (e)). As a special case of pool-based switching strategy, $T=0$ means that pool-based switching strategists will always act as pool excluders. The unique stable coexistent equilibrium point changes to a center surrounding by periodic closed orbits. The reason is that pool excluders invade defectors, cooperators invade pool excluders, and defectors invade cooperators (see Fig.\ref{fig3}(f)). To analyze the dynamics in the interior of $S_{3}$, we introduce a new variable $\varepsilon=\frac{x}{x+y}$, which represents the fraction of
cooperators among members who do not contribute to the exclusion pool. This yields
\begin{eqnarray*}
\dot{\varepsilon}&=&-\varepsilon(1-\varepsilon)(P_{D}-P_{C}).
\end{eqnarray*}
By substituting $x=\varepsilon(1-z)$ and $\bar{P}=x(P_{C}-P_{D})+(1-z)(P_{D}-P_{F})+P_{F}$ into $\dot{z}=z(P_{F}-\bar{P})$, we have $\dot{z}=z[x(P_{D}-P_{C})-(1-z)(P_{D}-P_{F})]$.
Thus we have
\begin{eqnarray*}
\left\{
\begin{aligned}
\dot{\varepsilon}&=-\varepsilon(1-\varepsilon)[(1-z)^{N-1}\frac{rc(N-1)}{N}-rc+c],\label{function3}\\
\dot{z}&=z(1-z)[rc-c-\delta-\tau-\varepsilon(rc-c)].
\end{aligned}
\right.
\end{eqnarray*}
Through detailed theoretical analysis, we can prove that the system is a conservative Hamiltonian system (see Appendix D.3 for details).

\begin{figure*}[!t]
\begin{center}
\includegraphics[width=15cm]{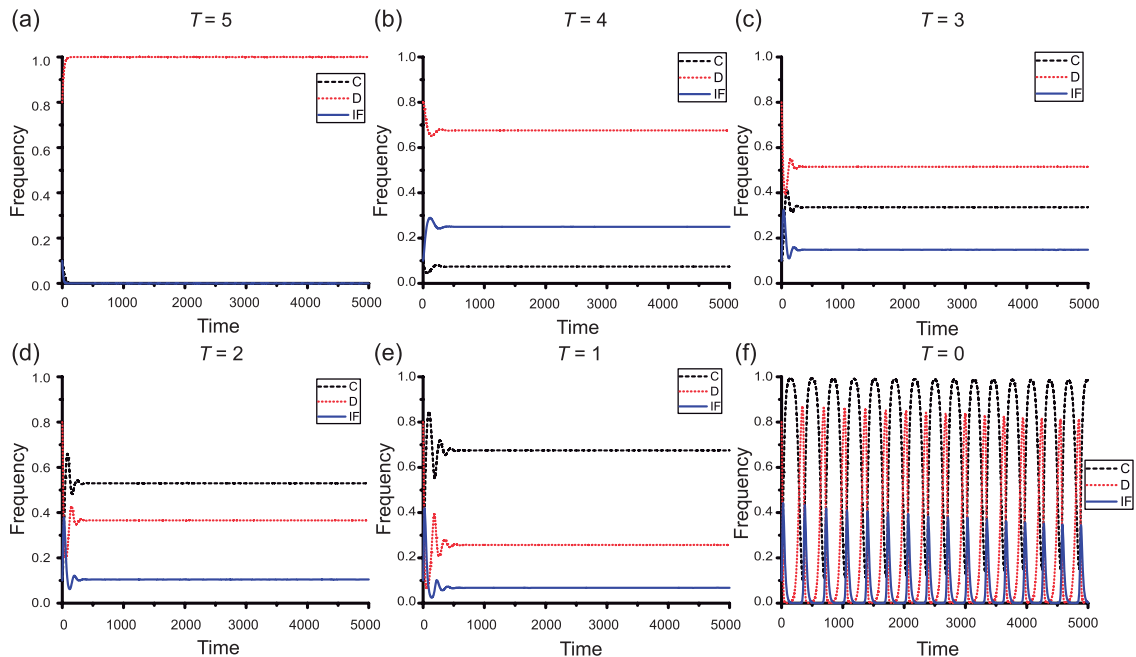}
\caption{Time evolution of the fractions of three strategies $C$ (black dash line), $D$ (red dot line), and $I_{F}$ (blue solid line) for different switching threshold values $T$. Initial conditions: $(x, y, z)=(0.1, 0.8, 0.1)$. Parameters: $N=5$, $r=3$,
$c=1$, $B=0.4$, $G=\delta=0.4$, $\tau=0.1$, and $c_{E}=0.4$. The threshold values are (a) $T=5$, (b) $T=4$, (c) $T=3$, (d) $T=2$, (e) $T=1$, and (f) $T=0$, respectively.}\label{fig4}
\end{center}
\end{figure*}

To further explain the results presented in Fig.\ref{fig3}, we continue by showing the time evolution of the frequencies of these three strategies for different values of $T$ in Fig.\ref{fig4}. When $T=N$, defection strategy can be dominant, which is irrelevant to the initial state (see Fig.\ref{fig4}(a)). Furthermore, by decreasing the tolerance level, however, we can observe the fraction of cooperators increases (see Fig.\ref{fig4}(b), (c), (d), and (e)). Of particular note is that pure cooperation strategy can become the most advantageous strategy when $T$ is set to an intermediate value, e.g. $T=2$. As $T$ is decreased further, the advantage of cooperators is further enhanced (see Fig.\ref{fig4}(d) and (e)).
Particularly, for $T=0$ the frequencies of $C, D$, and $I_{F}$ display periodic oscillations, which correspond to the limit cycle in Fig.\ref{fig3}(f).

\section{Discussion}

Thus far, many previous theoretical works have revealed that pro-social punishment and exclusion strategies can both maintain sufficiently high levels of public cooperation no matter whether these two strategies are implemented separately or jointly \cite{sasaki13,szolnokia17,liul17}. However, few studies have thus far considered the combined use of these two strategies despite it is particularly common in our real society. In this paper, we have introduced the switching strategy with which individuals can either punish or exclude free-riders in the public goods game (PGG), depending on the number of defectors in the group. Based on the evolutionary game theoretical models we have studied the evolutionary dynamics in the well-mixed population with two different switching forms, by focusing particularly on the role of switching threshold played in the evolutionary dynamics of cooperation.

We have shown that a stable coexistence state among cooperators, defectors, and peer-based switching strategists can appear when an intermediate switching threshold is used. In addition, the reduction of switching threshold can enhance the level of public cooperation.
It is necessary to point out that the special cases in our model ($T=N$ and $T=0$ correspond to peer exclusion and peer punishment, respectively) have been investigated in a recent work \cite{sasaki13}, which demonstrated that social exclusion strategy can not only avert free-riders, but also prevent second-order free-riders from invading, which solves the two substantial difficulties of peer punishment. Interestingly, we find that the introduction of an observation cost will completely change the evolutionary results of these two strategies in our model. Concretely, the coexistence of cooperators and peer punishers will disappear, and defection becomes a global stability strategy (see Fig.\ref{fig1}(a)). Or, the periodic oscillations among the three strategies replace the coexistence of cooperators and peer excluders (see Fig.\ref{fig1}(f)). Although similar shapes of oscillations have been presented in previous works \cite{Zschaler10_njp,reichenbach_06_pre} where oscillations are caused by the feedback between players' payoffs and their local interaction topology or the cyclic dominance of the three species, the mechanisms for the oscillations between our work and the two studies mentioned above are different. In our work, we investigate the evolutionary dynamics among cooperators, defectors, and switching strategists in an infinite population. The oscillation is caused by the mutual restriction among these three strategists, that is, defectors will be excluded by peer excluders, defectors defeat cooperators, and cooperators invade peer excluders.

We have also investigated the evolutionary dynamics of pool-based switching strategy and revealed that in addition to the two special cases ($T=0$ and $T=N$ correspond to the case of pool punishment and pool exclusion, respectively) mentioned above, the intermediate switching threshold values can guarantee the stable coexistence state of cooperators, defectors, and pool-based switching strategists. It needs to point that the result is stile valid even the cost of exclusion exceeds four times than we given in Fig.\ref{fig3}. Furthermore, the decrease of threshold value will weaken the advantage of defectors in evolution. Particularly, if switching threshold is zero, the result shows that there can be isolated periodic orbits and hence cooperators, defectors, and pool excluders can coexist by forming limit cycles (see Fig.\ref{fig3}(f)).

Finally, we have to note that the switching strategy proposed in this work can better induce the stable coexistence state for cooperation compared with pure punishment or pure exclusion strategies. Besides, such switching strategy can flexibly manage public goods in different environments, such as punishing exiguous free-riders in favorable surroundings or excluding massive free-riders in extremely unfavorable environments. Thus, the presently discussed strategy offers a simple, but still effective, way on how we can better promote the stable coexistence of different strategies including cooperation.




\section*{Appendix A}
\renewcommand{\theequation}{A.\arabic{equation}}

\subsection*{A.1 Peer punishment}\label{AppendixA.1}

We first study the replicator dynamics for defectors (D), cooperators (C), and peer punishers (W). This corresponds to the special case for peer-based switching strategy with $T=N$. We denote by $x, y$, and $z$ the frequencies of C, D, and W, respectively. Thus $x, y, z\geq0$ and $x+y+z=1$. The evolutionary fate of the population can be modeled by the replicator equations, given as
\begin{eqnarray}
\left\{
\begin{aligned}
\dot{x}&=x(P_{C}-\bar{P}),\\
\dot{y}&=y(P_{D}-\bar{P}),\\
\dot{z}&=z(P_{W}-\bar{P}),
\end{aligned}
\right.
\end{eqnarray}
where $P_{C}, P_{D}$, and $P_{W}$ denote the expected payoffs of these three strategies and $\bar{P}=xP_{C}+yP_{D}+zP_{W}$ describes the average payoff of the entire population.

Accordingly, the expected payoffs of these three strategies can be respectively given by
\begin{eqnarray*}
P_{C}&=&\frac{rc}{N}(N-1)(x+z)+\frac{rc}{N}-c,  \\
P_{D}&=&\frac{rc}{N}(N-1)(x+z)-(N-1)z\beta,  \\
P_{W}&=&\frac{rc}{N}(N-1)(x+z)+\frac{rc}{N}-c-(N-1)y\gamma-\tau.
\end{eqnarray*}

Since $P_{W}<P_{C}$, there is no interior fixed point.
Then we investigate the dynamics on each edge of the simplex
$S_{3}$. On the edge C-D we have $z=0$, resulting in
$\dot{y}=y(1-y)(P_{D}-P_{C})=y(1-y)(c-\frac{rc}{N})>0$. Thus the
direction of the dynamics goes from C to D.
On the edge D-W, since $x=0$ and $y+z=1$, we have
$\dot{z}=z(1-z)(P_{W}-P_{D})$. Here we assume that $0<c+\tau-\frac{rc}{N}+(N-1)\gamma<(N-1)(\gamma+\beta)$, thus solving $P_{W}=P_{D}$ results in $z=\frac{c+(N-1)\gamma+\tau-\frac{rc}{N}}{(N-1)(\beta+\gamma)}$, which means that there exists a boundary fixed point on the edge D-W.
On the edge C-W, since $y=0$ and $x+z=1$, we have
$\dot{x}=x(1-x)(P_{C}-P_{W})=x(1-x)\tau>0$, thus the
direction of the dynamics goes from W to C.

Therefore there are four equilibria, namely, three vertex fixed points ($(x, y, z)=(0, 0, 1), (1, 0, 0),$ and $(0, 1, 0)$) and the boundary fixed point ($(x, y, z)=(0, 1-\frac{c+(N-1)\gamma+\tau-\frac{rc}{N}}{(N-1)(\beta+\gamma)},\frac{c+(N-1)\gamma+\tau-\frac{rc}{N}}{(N-1)(\beta+\gamma)})$) in the simplex $S_3$.

\subsubsection*{A.2 The stabilities of equilibria}

Here we define
\begin{eqnarray*}
f(x,y)&=x[(1-x)(P_{C}-P_{W})-y(P_{D}-P_{W})],\\
g(x,y)&=y[(1-y)(P_{D}-P_{W})-x(P_{C}-P_{W})].
\end{eqnarray*}
Then the Jacobian matrix of the equation system is
\begin{equation}
J=\begin{bmatrix}
\frac{\partial{f(x,y)}}{\partial{x}} & \frac{\partial{f(x,y)}}{\partial{y}}\\
\frac{\partial{g(x,y)}}{\partial{x}} & \frac{\partial{g(x,y)}}{\partial{y}}
\end{bmatrix},
\end{equation}
where
\begin{eqnarray}
\left\{
\begin{aligned}
\frac{\partial{f(x,y)}}{\partial{x}}&=[(1-x)(P_{C}-P_{W})-y(P_{D}-P_{W})]\\
&+x[-(P_{C}-P_{W})+(1-x)\frac{\partial}{\partial{x}}(P_{C}-P_{W})\\
&-y\frac{\partial}{\partial{x}}(P_{D}-P_{W})],\\
\frac{\partial{f(x,y)}}{\partial{y}}&=x[(1-x)\frac{\partial}{\partial{y}}(P_{C}-P_{W})-(P_{D}-P_{W})\\
&-y\frac{\partial}{\partial{y}}(P_{D}-P_{W})],\\
\frac{\partial{g(x,y)}}{\partial{x}}&=y[(1-y)\frac{\partial}{\partial{x}}(P_{D}-P_{W})-(P_{C}-P_{W})\\
&-x\frac{\partial}{\partial{x}}(P_{C}-P_{W})],\\
\frac{\partial{g(x,y)}}{\partial{y}}&=[(1-y)(P_{D}-P_{W})-x(P_{C}-P_{W})]\\
&+y[-(P_{D}-P_{W})+(1-y)\frac{\partial}{\partial{y}}(P_{D}-P_{W})\\
&-x\frac{\partial}{\partial{y}}(P_{C}-P_{W})].
\end{aligned}
\right.
\end{eqnarray}

Then we have the following conclusion.

\noindent\textbf{Theorem 1.}
\emph{For $1<r<N$, only the fixed point $(0, 1, 0)$ is stable, and the other equilibria $(0, 0, 1), (1, 0, 0),$ and $(0, 1-\frac{c+(N-1)\gamma+\tau-\frac{rc}{N}}{(N-1)(\beta+\gamma)},\frac{c+(N-1)\gamma+\tau-\frac{rc}{N}}{(N-1)(\beta+\gamma)})$ are unstable.}

\emph{Proof.}
$(1)$ For $(x,y,z)=(0,0,1)$, the Jacobian is
\begin{equation}
J=\begin{bmatrix}
\tau & 0\\
0 & c+\tau-\frac{rc}{N}-(N-1)\beta
\end{bmatrix},
\end{equation}
thus the fixed point is unstable since $\tau>0$.\\
$(2)$ For $(x,y,z)=(1,0,0)$, the Jacobian is
\begin{equation}
J=\begin{bmatrix}
-\tau & -(c+\tau-\frac{rc}{N})\\
0 & c-\frac{rc}{N}
\end{bmatrix},
\end{equation}
thus the fixed point is unstable since $1-\frac{r}{N}>0$.\\
$(3)$ For $(x,y,z)=(0,1,0)$, the Jacobian is
\begin{equation}
J=\begin{bmatrix}
\frac{rc}{N}-c & 0\\
-\tau-(N-1)\gamma & \frac{rc}{N}-c-\tau-(N-1)\gamma
\end{bmatrix},
\end{equation}
thus the fixed point is stable since $\frac{rc}{N}-c<0$.\\
$(4)$ For $(x,y,z)=(0,1-\frac{c+(N-1)\gamma+\tau-\frac{rc}{N}}{(N-1)(\beta+\gamma)},\frac{c+(N-1)\gamma+\tau-\frac{rc}{N}}{(N-1)(\beta+\gamma)})$, the Jacobian is
\begin{equation}
J=\begin{bmatrix}
a_{11} & 0\\
a_{21} & a_{22}
\end{bmatrix},
\end{equation}
where $a_{11}=(N-1)y\gamma+\tau, a_{21}=y(1-y)\beta(N-1)-(N-1)y^{2}\gamma-\tau y$, and $a_{22}=y(1-y)(N-1)(\beta+\gamma)$, thus the fixed point is unstable since $y(1-y)(N-1)(\beta+\gamma)>0$ and $(N-1)y\gamma+\tau>0$.

\section*{Appendix B}\label{AppendixA.2}
\renewcommand{\theequation}{B.\arabic{equation}}
\subsection*{B.1 Peer exclusion}

We now consider another special case for peer-based switching strategy with $T=0$. Thus the replicator
equations are written as
\begin{eqnarray}
\left\{
\begin{aligned}
\dot{x}&=x(P_{C}-\bar{P}),\\
\dot{y}&=y(P_{D}-\bar{P}),\\
\dot{z}&=z(P_{E}-\bar{P}),
\end{aligned}
\right.\label{system2.3}
\end{eqnarray}
where $\bar{P}=xP_{C}+yP_{D}+zP_{E}$ represents the average payoff of the entire population.

We assume that exclusion never fails. In this condition, we can formalize the expected payoffs as follows
\begin{eqnarray}
P_{C}&=&rc-c-(1-z)^{N-1}\frac{rc(N-1)y}{N(1-z)},\\
P_{D}&=&(1-z)^{N-1}\frac{rc(N-1)x}{N(1-z)},\\
P_{E}&=&rc-c-(N-1)yc_{E}-\tau.
\end{eqnarray}

\noindent\textbf{Remark 1:} Because of $z=1-x-y$, the system (\ref{system2.3}) becomes
\begin{eqnarray}\label{system2}
\left\{
\begin{aligned}
\dot{x}&=x[(1-x)(P_{C}-P_{E})-y(P_{D}-P_{E})],\\
\dot{y}&=y[(1-y)(P_{D}-P_{E})-x(P_{C}-P_{E})],
\end{aligned}
\right.\label{2.2}
\end{eqnarray}
where
\begin{eqnarray*}
P_{C}-P_{E}&=&(N-1)yc_{E}+\tau-(1-z)^{N-1}\frac{rc(N-1)y}{N(1-z)},\\
P_{D}-P_{E}&=&(1-z)^{N-1}\frac{rc(N-1)x}{N(1-z)}\\
&-&rc+c+(N-1)yc_{E}+\tau.
\end{eqnarray*}

\noindent\textbf{Theorem 2\label{theorem2}}. \emph{For $[\frac{N(r-1)}{r(N-1)}]^{\frac{1}{N-1}}(N-1)c_{E}<rc-c-\tau<\min\{(N-1)c_{E}, [\frac{N(r-1)}{r(N-1)}]^{\frac{1}{N-1}}(N-1)c_{E}+\frac{rc-c}{[\frac{N(r-1)}{r(N-1)}]^{\frac{1}{N-1}}}-(N-1)c_{E}\}$, the system (\ref{system2}) has five fixed points, namely, $(x, y, z)=(0, 0, 1), (1, 0, 0), (0, 1, 0), (0, \frac{rc-c-\tau}{(N-1)c_{E}}, 1-\frac{rc-c-\tau}{(N-1)c_{E}}),$ and $([\frac{N(r-1)}{r(N-1)}]^{\frac{1}{N-1}}-\frac{\tau}{\frac{(r-1)c}{[\frac{N(r-1)}{r(N-1)}]^{\frac{1}{N-1}}}-(N-1)c_{E}}, \frac{\tau}{\frac{(r-1)c}{[\frac{N(r-1)}{r(N-1)}]^{\frac{1}{N-1}}}-(N-1)c_{E}}, 1-[\frac{N(r-1)}{r(N-1)}]^{\frac{1}{N-1}})$.}

\emph{Proof.}
By solving system equations (\ref{system2}), we can easily know that there are three vertex fixed points, namely, $(0, 0, 1), (0, 1, 0),$ and $(1, 0, 0)$. Then solving $P_{C}=P_{D}$ results in $z=1-[\frac{N(r-1)}{r(N-1)}]^{\frac{1}{N-1}}$. When $\frac{(r-1)c}{[\frac{N(r-1)}{r(N-1)}]^{\frac{1}{N-1}}}-(N-1)c_{E}>\tau$, solving $P_{C}=P_{E}$ leads to $y=\frac{\tau}{\frac{(r-1)c}{[\frac{N(r-1)}{r(N-1)}]^{\frac{1}{N-1}}}-(N-1)c_{E}}$. Thus there exists an interior fixed point
$([\frac{N(r-1)}{r(N-1)}]^{\frac{1}{N-1}}-\frac{\tau}{\frac{(r-1)c}{[\frac{N(r-1)}{r(N-1)}]^{\frac{1}{N-1}}}-(N-1)c_{E}}, \frac{\tau}{\frac{(r-1)c}{[\frac{N(r-1)}{r(N-1)}]^{\frac{1}{N-1}}}-(N-1)c_{E}}, 1-[\frac{N(r-1)}{r(N-1)}]^{\frac{1}{N-1}})$ in the simplex $S_{3}$.

Then, we study the dynamics on each edge of simplex $S_{3}$.
On the edge E-D, $y+z=1$ results in $\dot{z}=z(1-z)(P_{E}-P_{D})=z(1-z)[rc-c-\tau-(N-1)yc_{E}]$, thus there is an equilibrium $y=\frac{rc-c-\tau}{(N-1)c_{E}}$ for $0<rc-c-\tau<(N-1)c_{E}$, otherwise E can perform better than D. On the edge C-E, we have $x+z=1$ and $\dot{x}=x(1-x)(P_{C}-P_{E})=x(1-x)\tau>0$, thus the
direction of the dynamics goes from E to C. On the edge C-D, D can defeat C, as presented in Appendix A.1.

\subsection*{B.2 The stabilities of equilibria}\label{AppendixA.21}

\noindent\textbf{Theorem 3.} \emph{In the conditions of Theorem 2, only the fixed point $(0, 1, 0)$ is stable, and the other equilibria $(0, 0, 1), (1, 0, 0), (0, \frac{rc-c-\tau}{(N-1)c_{E}}, 1-\frac{rc-c-\tau}{(N-1)c_{E}}),$ and $([\frac{N(r-1)}{r(N-1)}]^{\frac{1}{N-1}}-\frac{\tau}{\frac{(r-1)c}{[\frac{N(r-1)}{r(N-1)}]^{\frac{1}{N-1}}}-(N-1)c_{E}}, \frac{\tau}{\frac{(r-1)c}{[\frac{N(r-1)}{r(N-1)}]^{\frac{1}{N-1}}}-(N-1)c_{E}}, 1-[\frac{N(r-1)}{r(N-1)}]^{\frac{1}{N-1}})$ are unstable.\label{theorem3}}

\emph{Proof.}
We also use the Jacobian matrix of the system to study the stability of equilibria.\\
$(1)$ For $(x,y,z)=(0,0,1)$, the Jacobian is
\begin{equation}
J=\begin{bmatrix}
\tau & 0\\
0 & -(rc-c-\tau)
\end{bmatrix},
\end{equation}
thus the fixed point is unstable since $\tau>0$.\\
$(2)$ For $(x,y,z)=(1,0,0)$, the Jacobian is
\begin{equation}
J=\begin{bmatrix}
-\tau & -(c+\tau-\frac{rc}{N})\\
0 & c-\frac{rc}{N}
\end{bmatrix},
\end{equation}
thus the fixed point is a unstable since $1-\frac{r}{N}>0$.\\
$(3)$ For $(x,y,z)=(0,1,0)$, the Jacobian is
\begin{equation}
J=\begin{bmatrix}
\frac{rc}{N}-c & 0\\
a_{21} & rc-c-\tau-(N-1)c_{E}
\end{bmatrix},
\end{equation}
where $a_{21}=-(\frac{rc}{N}-rc+(N-1)c_{E}+\tau)$, thus the fixed point is unstable when $rc-c-\tau-(N-1)c_{E}>0$, while when $rc-c-\tau-(N-1)c_{E}<0$ it is stable. Particularly, when $rc-c-\tau-(N-1)c_{E}=0$, we can prove this fixed point is stable by using center manifold theorem (see Theorem. 4 for detail analysis).\\
$(4)$ When $rc-c-\tau-(N-1)c_{E}<0$ there is a boundary equilibrium point $(x,y,z)=(0, \frac{rc-c-\tau}{(N-1)c_{E}}, 1-\frac{rc-c-\tau}{(N-1)c_{E}})$, then the Jacobian is
\begin{equation}
J=\begin{bmatrix}
a_{11} & 0\\
a_{21} & a_{22}
\end{bmatrix},
\end{equation}
where $a_{11}=(N-1)yc_{E}-y^{N-1}\frac{rc(N-1)}{N}+\tau, a_{21}=y^{N-1}\frac{rc}{N}(N-1)-(N-1)y^{2}c_{E}-y\tau,$ and $a_{22}=y(1-y)(N-1)c_{E}$, thus the fixed point is unstable since $y(1-y)(N-1)c_{E}>0$.\\
$(5)$ When $\tau<rc-c-[\frac{N(r-1)}{r(N-1)}]^{\frac{1}{N-1}}(N-1)c_{E}<\tau+\frac{rc-c}{[\frac{N(r-1)}{r(N-1)}]^{\frac{1}{N-1}}}-(N-1)c_{E}$, there is an interior equilibrium point $(x,y,z)=([\frac{N(r-1)}{r(N-1)}]^{\frac{1}{N-1}}-\frac{\tau}{\frac{(r-1)c}{[\frac{N(r-1)}{r(N-1)}]^{\frac{1}{N-1}}}-(N-1)c_{E}}, \\ \frac{\tau}{\frac{(r-1)c}{[\frac{N(r-1)}{r(N-1)}]^{\frac{1}{N-1}}}-(N-1)c_{E}}, 1-[\frac{N(r-1)}{r(N-1)}]^{\frac{1}{N-1}})$, we define the equilibrium point as $(x^{*}, y^{*}, z^{*})$ hereafter. And the elements in the
Jacobian matrix for this equilibrium point are written as
\setlength{\arraycolsep}{0.0em}
\begin{eqnarray*}
\left\{
\begin{aligned}
\frac{\partial{f}}{\partial{x}}(x^{*},y^{*})&=x^{*}[(1-x^{*})\frac{\partial}{\partial{x}}(P_{C}-P_{E})-y^{*}\frac{\partial}{\partial{x}}(P_{D}-P_{E})],\\
\frac{\partial{f}}{\partial{y}}(x^{*},y^{*})&=x^{*}[(1-x^{*})\frac{\partial}{\partial{y}}(P_{C}-P_{E})-y^{*}\frac{\partial}{\partial{y}}(P_{D}-P_{E})],\\
\frac{\partial{g}}{\partial{x}}(x^{*},y^{*})&=y^{*}[(1-y^{*})\frac{\partial}{\partial{x}}(P_{D}-P_{E})-x^{*}\frac{\partial}{\partial{x}}(P_{C}-P_{E})],\\
\frac{\partial{g}}{\partial{y}}(x^{*},y^{*})&=y^{*}[(1-y^{*})\frac{\partial}{\partial{y}}(P_{D}-P_{E})-x^{*}\frac{\partial}{\partial{y}}(P_{C}-P_{E})],
\end{aligned}
\right.
\end{eqnarray*}
\setlength{\arraycolsep}{5.0pt}
where
\setlength{\arraycolsep}{0.0em}
\begin{eqnarray*}
\left\{
\begin{aligned}
\frac{\partial}{\partial{x}}(P_{C}-P_{E})&=-(x^{*}+y^{*})^{N-3}\frac{rcy^{*}(N-1)(N-2)}{N},\\
\frac{\partial}{\partial{y}}(P_{C}-P_{E})&=(N-1)c_{E}-(x^{*}+y^{*})^{N-3}\frac{rc(N-1)}{N}\\
&[(N-1)y^{*}+x^{*}],\\
\frac{\partial}{\partial{x}}(P_{D}-P_{E})&=(x^{*}+y^{*})^{N-3}\frac{rc(N-1)}{N}[(N-1)x^{*}+y^{*}],\\
\frac{\partial}{\partial{y}}(P_{D}-P_{E})&=(N-1)c_{E}+(x^{*}+y^{*})^{N-3}\\
&\frac{rcx^{*}(N-1)(N-2)}{N}.
\end{aligned}
\right.
\end{eqnarray*}
\setlength{\arraycolsep}{5.0pt}
Then we define that $p_{1}=\frac{\partial{f}}{\partial{x}}(x^{*},y^{*})\frac{\partial{g}}{\partial{y}}(x^{*},y^{*})-\frac{\partial{f}}{\partial{y}}(x^{*},y^{*})\frac{\partial{g}}{\partial{x}}(x^{*},y^{*})$ and $q_{1}=\frac{\partial{f}}{\partial{x}}(x^{*},y^{*})+\frac{\partial{g}}{\partial{y}}(x^{*},y^{*})$. We know that
\begin{eqnarray*}
p_{1}&=&x^{*}y^{*}(1-x^{*}-y^{*})[\frac{\partial}{\partial{x}}(P_{C}-P_{E})\frac{\partial}{\partial{y}}(P_{D}-P_{E})\\
&-&\frac{\partial}{\partial{y}}(P_{C}-P_{E})\frac{\partial}{\partial{x}}(P_{D}-P_{E})]\nonumber\\
&=&x^{*}y^{*}(1-x^{*}-y^{*})\frac{rc(N-1)^{3}}{N}(x^{*}+y^{*})^{N-2}[(x^{*}\\
&+&y^{*})^{N-2}\frac{rc}{N}-1]\nonumber\\
&<&0,
\end{eqnarray*}
and
\begin{eqnarray*}
q_{1}&=&x^{*}(1-x^{*})\frac{\partial}{\partial{x}}(P_{C}-P_{E})+y^{*}(1-y^{*})\frac{\partial}{\partial{y}}(P_{D}-P_{E})\nonumber\\
&-&x^{*}y^{*}[\frac{\partial}{\partial{x}}(P_{D}-P_{E})+\frac{\partial}{\partial{y}}(P_{C}-P_{E})]\nonumber\\
&=&(N-1)c_{E}y^{*}(1-y^{*}-x^{*})\nonumber\\
&>&0.
\end{eqnarray*}

We thus conclude that the Jacobian matrix has a positive eigenvalue. Therefore the interior equilibrium point is unstable.

\noindent\textbf{Theorem 4.} \emph{For $rc-c-\tau-(N-1)c_{E}=0$, the equilibrium point $(0, 1, 0)$} is stable.\label{Theorem 4}

\emph{Proof.} Because $y = 1 - x - z$, the dynamic equations (\ref{system2}) become
\begin{eqnarray}\label{system2.2}
\left\{
\begin{aligned}
\dot{x}&=x[(1-x)(P_{C}-P_{D})-z(P_{E}-P_{D})],\\
\dot{z}&=z[(1-z)(P_{E}-P_{D})-x(P_{C}-P_{D})],
\end{aligned}
\right.\label{2.3}
\end{eqnarray}
where
\begin{eqnarray*}
P_{C}-P_{D}&=&rc-c-(1-z)^{N-1}\frac{rc(N-1)}{N},\\
P_{E}-P_{D}&=&rc-c-(N-1)yc_{E}-\tau\\
&-&(1-z)^{N-1}\frac{rc(N-1)x}{N(1-z)}.
\end{eqnarray*}

We know that $(x, z)=(0, 0)$ is equilibrium point of system (\ref{system2.2}). Then the Jacobian is
\begin{equation}
A=\begin{bmatrix}
\frac{rc}{N}-c & 0\\
0 & rc-c-\tau-(N-1)c_{E}
\end{bmatrix}.
\end{equation}
When $rc-c-\tau-(N-1)c_{E}=0$, we know that the eigenvalues of the Jacobian for the fixed point $(x, z)=(0, 0)$ are $0$ and $\frac{rc}{N}-c$. In this condition, we study the stability of the equilibrium point by further using the center manifold
theorem \cite{Khalil96,carr81,Hassard81}. To do that, we construct a matrix $M$, whose column elements are the eigenvectors of the matrix $A$, given as
\begin{equation}
M=\begin{bmatrix}
0 & 1\\
1 & 0
\end{bmatrix}.
\end{equation}
Let $T=M^{-1}$, then we have
\begin{equation}
TAT^{-1}=\begin{bmatrix}
0 & 0\\
0 & \frac{rc}{N}-c
\end{bmatrix}.
\end{equation}
Using variable substitution, we have
\begin{equation}
\begin{bmatrix}
j \\
i
\end{bmatrix}=T\begin{bmatrix}
x\\
z
\end{bmatrix}=\begin{bmatrix}
0 & 1\\
1 & 0
\end{bmatrix}\begin{bmatrix}
x\\
z
\end{bmatrix}=\begin{bmatrix}
z\\
x
\end{bmatrix}.
\end{equation}
Therefore, the system (\ref{2.3}) can be rewritten as
\begin{eqnarray}
\left\{
\begin{aligned}
\dot{j}&=j(1-j)[rc-c-(N-1)(1-j-i)c_{E}\\
&-\tau-(1-j)^{N-1}\frac{rc(N-1)i}{N(1-j)}]\nonumber\\
&-ji[rc-c-(1-j)^{N-1}\frac{rc(N-1)}{N}],\\
\dot{i}&=i(1-i)[rc-c-(1-j)^{N-1}\frac{rc(N-1)}{N}]\nonumber\\
&-ji[rc-c-(N-1)(1-j-i)c_{E}-\tau\\
&-(1-j)^{N-1}\frac{rc(N-1)i}{N(1-j)}].
\end{aligned}
\right.
\end{eqnarray}
Using the center manifold theorem, we have that $j = h(i)$ is a center manifold for the above system.
Then the dynamics on the center manifold can be described by
\begin{eqnarray}\label{A35}
\dot{i}&=&i(1-i)[rc-c-(1-h(i))^{N-1}\frac{rc(N-1)}{N}]\nonumber\\
&-&h(i)i[rc-c-(N-1)(1-h(i)-i)c_{E}-\tau\nonumber\\
&-&(1-h(i))^{N-1}\frac{rc(N-1)i}{N(1-h(i))}].
\end{eqnarray}
We start to try $h(i)=O(i^2)$, thus the system (\ref{A35}) can be expressed as
\begin{eqnarray}
\dot{i}=i(1-i)[rc-c-\frac{rc(N-1)}{N}]+O(|i|^{3}).
\end{eqnarray}
By defining $m(i)=i(1-i)[rc-c-\frac{rc(N-1)}{N}]$, we have $m^{'}(i)=(1-2i)[rc-c-\frac{rc(N-1)}{N}]$. Since $m^{'}(0)<0$,
thus we can judge that $i=0$ is asymptotically stable. Therefore, we can know the fixed point $(i, h(i))=(0, 0)$ is also stable for the system (\ref{system2.2}). Accordingly, the fixed point $(0, 1, 0)$ is stable as well \cite{Khalil96,carr81,Hassard81}.

\subsection*{B.3 Heteroclinic cycle}\label{AppendixA.22}

In this subsection, we show that there is a stable heteroclinic cycle on the boundary of the simplex $CDI_{E}$ (see Fig. \ref{fig1} (f)). When $rc-c-\tau-(N-1)c_{E}>0$ and $r<N$, we know that the three vertex equilibrium points ($C, D$, and $I_{E}$) are all saddle nodes, and the heteroclinic trajectories can display on the three edges ($CD, DI_{E}$, and $I_{E}C$). All of these guarantee the existence of the heteroclinic cycle on the boundary $S_{3}$. In the following, we will prove that the heteroclinic cycle is asymptotically stable.

According to Theorem 3, we can get the eigenvalues of the Jacobian matrix of the three vertex equilibrium points, namely, $\lambda_{I_{E}}^{-}=-(rc-c-\tau), \lambda_{I_{E}}^{+}=\tau, \lambda_{C}^{-}=-\tau, \lambda_{C}^{+}=c-\frac{rc}{N}, \lambda_{D}^{-}=\frac{rc}{N}-c,$ and $\lambda_{D}^{+}=rc-c-\tau-(N-1)c_{E}$. Then we define that $\lambda_{I_{E}}=-\frac{\lambda_{I_{E}}^{-}}{\lambda_{I_{E}}^{+}}, \lambda_{C}=-\frac{\lambda_{C}^{-}}{\lambda_{C}^{+}},$ and $\lambda_{D}=-\frac{\lambda_{D}^{-}}{\lambda_{D}^{+}}$, and we have $\lambda=\lambda_{I_{E}}\lambda_{C}\lambda_{D}=\frac{rc-c-\tau}{rc-c-\tau-(N-1)c_{E}}>1$. Thus the heteroclinic cycle is asymptotically stable \cite{park18}.

\subsection*{B.4 Limit cycle}\label{AppendixA.23}

Next, we prove that a stable limit cycle can exist in the simplex $CDI_{E}$ (see Fig. \ref{fig1} (f)). According to the above theoretical analysis, we know that the interior fixed point of the system is unstable. Now we set a closed domain $\Gamma$ that contains the interior fixed point on the phase plane of the system (\ref{system2}).
Here we take a straight line $l=x+y-b=0$, where $b$ is an undetermined constant. In order to determine the direction of the trajectory of the system, we solve the derivative of the equation $l=x+y-b$. Thus we have
\begin{eqnarray}
\frac{\partial{l}}{\partial{t}}&=&\frac{\partial{x}}{\partial{t}}+\frac{\partial{y}}{\partial{t}}\nonumber\\
&=&x(1-x-y)[(N-1)yc_{E}+\tau-(x+y)^{N-1}\frac{rc(N-1)y}{N(x+y)}]\nonumber\\
&+&y(1-x-y)[(x+y)^{N-1}\frac{rc(N-1)x}{N(x+y)}-rc+c\nonumber\\
&+&(N-1)yc_{E}+\tau]\nonumber\\
&=&(1-x-y)[(N-1)yc_{E}(x+y)+(x+y)\tau-(rc-c)y]\nonumber\\
&=&(1-b)[(N-1)yc_{E}b+b\tau-(rc-c)y].
\end{eqnarray}

Thus when $b$ is sufficiently large, we know that $\frac{\partial{l}}{\partial{t}}<0$. In this case, the straight line $l=x+y-b=0, x=0,$ and $y=0$ are enclosed in a closed domain $\Gamma$ that points to the interior of the boundary.

On the straight line $x=x^{*}$, we select a point $(x^{*}, y_{1})$, where $y_{1}$ is slightly larger than $y^{*}$, and the trajectory that passes through $(x^{*}, y_{1})$ surrounds the interior fixed point $(x^{*}, y^{*})$, then intersects with the straight line $x=x^{*}$ at another point $(x^{*}, y_{2})$ which meets $y_{2}>y^{*}$. Since the interior fixed point $(x^{*}, y^{*})$ is unstable, we have $y_{2}>y_{1}$. The trajectory from $(x^{*}, y_{1})$ to $(x^{*}, y_{2})$ and the line segment $\overline{y_{1}y_{2}}$ form a closed domain $\Gamma_{0}$ containing the interior fixed point. The trajectories on the boundary of the closed domain $\Gamma_{0}$ are all diverged outward. Thus a ring domain is formed between the boundaries of the closed domains $\Gamma$ and $\Gamma_{0}$, and the trajectories on the boundary of the inner and outer ring will go to the interior of the domain. Accordingly, we prove that there exist a stable limit cycle in the closed domain $\Gamma$ \cite{Khalil96}.

\section*{Appendix C}
\renewcommand{\theequation}{C.\arabic{equation}}
\subsection*{C.1 Pool punishment}\label{AppendixA.3}

Next, we analyze the replicator dynamics for public goods game with pool punishment, which corresponds to the special case for pool-based switching strategy with $T=N$. Accordingly, there are three strategists, namely, cooperators, defectors, and pool punishers, respectively. Thus the replicator equations can be written as
\begin{eqnarray}\label{system3.1}
\left\{
\begin{aligned}
\dot{x}=x(P_{C}-\bar{P}),\\
\dot{y}=y(P_{D}-\bar{P}),\\
\dot{z}=z(P_{V}-\bar{P}),
\end{aligned}
\right.
\end{eqnarray}
where $\bar{P}=xP_{C}+yP_{D}+zP_{V}$ is the average payoff of the whole population. Then the expected payoffs of these three strategies can be respectively given by
\begin{eqnarray}
P_{C}&=&\frac{rc}{N}(N-1)(x+z)+\frac{rc}{N}-c,  \\
P_{D}&=&\frac{rc}{N}(N-1)(x+z)-B(N-1)z,  \\
P_{V}&=&\frac{rc}{N}(N-1)(x+z)+\frac{rc}{N}-c-G-\tau,
\end{eqnarray}
where $(N-1)(x+z)$ denotes the expected number of contributors among the $N-1$ co-players, and $B(N-1)z$ represents the expected fine on a defector.

\noindent\textbf{Theorem 5\label{theorem5}.} \emph{For $r<N$, the system (\ref{system3.1}) has four fixed points, namely, $(x, y, z)=(0, 0, 1), (1, 0, 0), (0, 1, 0),$ and $(0, 1-\frac{N(c+G+\tau)-rc}{N(N-1)B}, \frac{N(c+G+\tau)-rc}{N(N-1)B})$.}

\emph{Proof.} By solving the system equations (\ref{system3.1}), we can know that there are three vertex equilibrium points, namely, $(x, y, z) = (0, 0, 1), (0, 1, 0)$, and $(1, 0, 0)$.

There is no interior fixed point since $P_{V}<P_{C}$. Then, we study the dynamics on each edge of simplex $S_{3}$.
On the edge V-D, $y+z=1$ results in $\dot{z}=z(1-z)(P_{V}-P_{D})=z(1-z)[\frac{rc}{N}-c-G-\tau+B(N-1)z]$, thus there is an equilibrium $z=\frac{N(c+G+\tau)-rc}{N(N-1)B}$ for $0<\frac{N(c+G+\tau)-rc}{N(N-1)B}<1$. On the edge C-V, we have $x+z=1$ and $\dot{x}=x(1-x)(P_{C}-P_{V})=x(1-x)(G+\tau)>0$, thus the
direction of the dynamics goes from V to C. On the edge C-D, D can defeat C, as presented in Appendix A.1.

\subsection*{C.2 The stabilities of equilibria}

\textbf{Theorem 6\label{theorem6}.} \emph{In the conditions of Theorem 5, the fixed point (0, 1, 0) is stable, while the others ($(0,0,1), (1,0,0),$ and $(0, 1-\frac{N(c+G+\tau)-rc}{N(N-1)B}, \frac{N(c+G+\tau)-rc}{N(N-1)B}))$ are unstable.}

\emph{Proof.}
$(1)$ For $(x,y,z)=(0,0,1)$, the Jacobian is
\begin{equation}
J=\begin{bmatrix}
G+\tau & 0\\
0 & -B(N-1)+c+G+\tau-\frac{rc}{N}
\end{bmatrix},
\end{equation}
thus the fixed point is unstable since $G+\tau>0$.\\
$(2)$ For $(x,y,z)=(1,0,0)$, the Jacobian is
\begin{equation}
J=\begin{bmatrix}
-\tau-G & -(c+\tau+G-\frac{rc}{N})\\
0 & c-\frac{rc}{N}
\end{bmatrix},
\end{equation}
thus the fixed point is unstable since $c-\frac{rc}{N}>0$.\\
$(3)$ For $(x,y,z)=(0,1,0)$, the Jacobian is
\begin{equation}
J=\begin{bmatrix}
\frac{rc}{N}-c & 0\\
-(G+\tau) & \frac{rc}{N}-c-\tau-G
\end{bmatrix},
\end{equation}
thus the fixed point is stable since $\frac{rc}{N}-c<0$.\\
$(4)$ For $(x,y,z)=(0, 1-\frac{N(c+G+\tau)-rc}{N(N-1)B}, \frac{N(c+G+\tau)-rc}{N(N-1)B})$, then the Jacobian is
\begin{equation}
J=\begin{bmatrix}
G+\tau & 0\\
y(1-y)B(N-1)-yGy\tau & y(1-y)(N-1)B
\end{bmatrix},
\end{equation}
thus the fixed point is unstable since $G+\tau>0$.

\section*{Appendix D}
\renewcommand{\theequation}{D.\arabic{equation}}
\subsection*{D.1 Pool exclusion}\label{AppendixA.4}

In this subsection, we study the evolutionary dynamics of pool exclusion, which corresponds to the special case for pool-base switching strategy with $T=0$.
Then the replicator equations become
\begin{eqnarray}\label{system4.1}
\left\{
\begin{aligned}
\dot{x}=x(P_{C}-\bar{P}),\\
\dot{y}=y(P_{D}-\bar{P}),\\
\dot{z}=z(P_{F}-\bar{P}),
\end{aligned}
\right.
\end{eqnarray}
where $P_{C}, P_{D}$, and $P_{F}$ are the expected payoffs of cooperators, defectors, and pool excluders, respectively.

We also assume that exclusion never fails. In this condition, we give the expected payoffs as follows
\begin{eqnarray}
P_{C}&=&rc-c-(1-z)^{N-1}\frac{rc(N-1)y}{N(1-z)},  \\
P_{D}&=&(1-z)^{N-1}\frac{rc}{N}(N-1)\frac{x}{1-z},  \\
P_{F}&=&rc-c-\delta-\tau.
\end{eqnarray}

\noindent\textbf{Theorem 7\label{theorem7}.} \emph{For $r<N$ and $\delta+\tau<rc-c$, the system (\ref{system4.1}) has four fixed points, namely, $(x,y,z)=(0, 0, 1), (1, 0, 0),$ \\$(0, 1, 0),$ and $([\frac{N(r-1)}{r(N-1)}]^{\frac{1}{N-1}}-\frac{(\delta+\tau)[\frac{N(r-1)}{r(N-1)}]^{\frac{1}{N-1}}}{(r-1)c},
\frac{(\delta+\tau)[\frac{N(r-1)}{r(N-1)}]^{\frac{1}{N-1}}}{(r-1)c},
1-[\frac{N(r-1)}{r(N-1)}]^{\frac{1}{N-1}})$.}

\emph{Proof.} By solving the system equations (\ref{system4.1}), we can know that there are three vertex equilibrium points, namely, $(0, 0, 1), (0, 1, 0)$, and $(1, 0, 0)$.

Solving $P_{C}=P_{D}$ results in
$z=1-[\frac{N(r-1)}{r(N-1)}]^{\frac{1}{N-1}}$.
Similarly, by solving $P_{C}=P_{F}$, we have
$y=\frac{(\delta+\tau)[\frac{N(r-1)}{r(N-1)}]^{\frac{1}{N-1}}}{(r-1)c}$. Thus when $\delta+\tau<rc-c$, there exists an interior fixed point $([\frac{N(r-1)}{r(N-1)}]^{\frac{1}{N-1}}-\frac{(\delta+\tau)[\frac{N(r-1)}{r(N-1)}]^{\frac{1}{N-1}}}{(r-1)c},
\frac{(\delta+\tau)[\frac{N(r-1)}{r(N-1)}]^{\frac{1}{N-1}}}{(r-1)c},
1-[\frac{N(r-1)}{r(N-1)}]^{\frac{1}{N-1}})$.

Then we investigate the dynamics on each edge of the simplex
$S_{3}$. On the edge C-D, we have $z=0$, resulting in
$\dot{y}=y(1-y)(P_{D}-P_{C})=y(1-y)(c-\frac{rc}{N})>0$. Thus the
direction of the dynamics goes from C to D.

On the edge D-F, we have
$\dot{z}=z(1-z)(P_{F}-P_{D})=z(1-z)(rc-c-\delta-\tau)>0$, thus the
direction of the dynamics goes from D to EC.

On the edge C-F, we have
$\dot{x}=x(1-x)(P_{C}-P_{F})=x(1-x)(\delta+\tau)>0$, thus the
direction of the dynamics goes from F to C.

\subsection*{D.2 The stabilities of equilibria}\label{AppendixA.41}

\noindent\textbf{Theorem 8\label{theorem8}.} \emph{In the conditions of Theorem. 7, the three vertex equilibria are unstable, and the interior fixed point is neutrally stable surrounded by closed and periodic orbits.}

\emph{Proof.}
$(1)$ For $(x,y,z)=(0,0,1)$, the Jacobian is
\begin{equation}
J(0,0,1)=\begin{bmatrix}
\delta+\tau & 0\\
0 & -rc+c+\delta+\tau
\end{bmatrix},
\end{equation}
thus the fixed point is unstable since $\delta+\tau>0$.\\
$(2)$ For $(x,y,z)=(1,0,0)$, the Jacobian is
\begin{equation}
J(1,0,0)=\begin{bmatrix}
-\delta-\tau & -(c+\delta+\tau-\frac{rc}{N})\\
0 & c-\frac{rc}{N}
\end{bmatrix},
\end{equation}
thus the fixed point is unstable since $1-\frac{r}{N}>0$.\\
$(3)$ For $(x,y,z)=(0,1,0)$, the Jacobian is
\begin{equation}
J(0,1,0)=\begin{bmatrix}
\frac{rc}{N}-c & 0\\
-(\frac{rc}{N}-rc+\delta+\tau) & rc-c-\delta-\tau
\end{bmatrix},
\end{equation}
thus the fixed point is unstable since $rc-c-\delta-\tau>0$.\\
$(4)$ For
$(x, y, z)=([\frac{N(r-1)}{r(N-1)}]^{\frac{1}{N-1}}-\frac{(\delta+\tau)[\frac{N(r-1)}{r(N-1)}]^{\frac{1}{N-1}}}{(r-1)c},
\frac{(\delta+\tau)[\frac{N(r-1)}{r(N-1)}]^{\frac{1}{N-1}}}{(r-1)c},
1-[\frac{N(r-1)}{r(N-1)}]^{\frac{1}{N-1}})$, we define the equilibrium point as $(x^{**}, y^{**}, z^{**})$ hereafter, and the elements in the
Jacobian matrix are written as
\setlength{\arraycolsep}{0.0em}
\begin{eqnarray*}
\left\{
\begin{aligned}
\frac{\partial{f}}{\partial{x}}(x^{**},y^{**})=x^{**}[(1-x^{**})\frac{\partial}{\partial{x}}(P_{C}-P_{F})\\
-y^{**}\frac{\partial}{\partial{x}}(P_{D}-P_{F})],\\
\frac{\partial{f}}{\partial{y}}(x^{**},y^{**})=x^{**}[(1-x^{**})\frac{\partial}{\partial{y}}(P_{C}-P_{F})\\
-y^{**}\frac{\partial}{\partial{y}}(P_{D}-P_{F})],\\
\frac{\partial{g}}{\partial{x}}(x^{**},y^{**})=y^{**}[(1-y^{**})\frac{\partial}{\partial{x}}(P_{D}-P_{F})\\
-x^{**}\frac{\partial}{\partial{x}}(P_{C}-P_{F})],\\
\frac{\partial{g}}{\partial{y}}(x^{**},y^{**})=y^{**}[(1-y^{**})\frac{\partial}{\partial{y}}(P_{D}-P_{F})\\
-x^{**}\frac{\partial}{\partial{y}}(P_{C}-P_{F})],
\end{aligned}
\right.
\end{eqnarray*}
\setlength{\arraycolsep}{5pt}
where
\setlength{\arraycolsep}{0.0em}
\begin{eqnarray*}
\left\{
\begin{aligned}
\frac{\partial}{\partial{x}}(P_{C}-P_{F})&=-(x^{**}+y^{**})^{N-3}\frac{rcy^{**}(N-1)(N-2)}{N},\\
\frac{\partial}{\partial{y}}(P_{C}-P_{F})&=-(x^{**}+y^{**})^{N-3}\frac{rc(N-1)}{N}\\
&[(N-1)y^{**}+x^{**}],\\
\frac{\partial}{\partial{x}}(P_{D}-P_{F})&=(x^{**}+y^{**})^{N-3}\frac{rc(N-1)}{N}\\
&[(N-1)x^{**}+y^{**}],\\
\frac{\partial}{\partial{y}}(P_{D}-P_{F})&=(x^{**}+y^{**})^{N-3}\frac{rcx^{**}(N-1)(N-2)}{N}.
\end{aligned}
\right.
\end{eqnarray*}
\setlength{\arraycolsep}{5pt}
Then we define that $p_{2}=\frac{\partial{f}}{\partial{x}}(x^{**},y^{**})\frac{\partial{g}}{\partial{y}}(x^{**},y^{**})-\frac{\partial{f}}{\partial{y}}(x^{**},y^{**})\frac{\partial{g}}{\partial{x}}(x^{**},y^{**})$ and $q_{2}=\frac{\partial{f}}{\partial{x}}(x^{**},y^{**})+\frac{\partial{g}}{\partial{y}}(x^{**},y^{**})$. Thus we know that
\setlength{\arraycolsep}{0.0em}
\begin{eqnarray*}
p_{2}&=&x^{**}y^{**}(1-x^{**}-y^{**})[\frac{\partial}{\partial{x}}(P_{C}-P_{F})\frac{\partial}{\partial{y}}(P_{D}-P_{F})\\
&-&\frac{\partial}{\partial{y}}(P_{C}-P_{F})\frac{\partial}{\partial{x}}(P_{D}-P_{F})]\nonumber\\
&=&x^{**}y^{**}(1-x^{**}-y^{**})(x^{**}+y^{**})^{2N-4}\frac{r^{2}c^{2}(N-1)^{3}}{N^{2}}\nonumber\\
&>&0,
\end{eqnarray*}
\setlength{\arraycolsep}{5pt}
and
\setlength{\arraycolsep}{0.0em}
\begin{eqnarray*}
q_{2}&=&x^{**}(1-x^{**})\frac{\partial}{\partial{x}}(P_{C}-P_{F})+y^{**}(1-y^{**})\frac{\partial}{\partial{y}}(P_{D}-P_{F})\nonumber\\
&-&x^{**}y^{**}[\frac{\partial}{\partial{y}}(P_{D}-P_{F})+\frac{\partial}{\partial{y}}(P_{C}-P_{F})]\nonumber\\
&=&x^{**}y^{**}(x^{**}+y^{**})^{N-3}\frac{rc(N-1)(N-2)}{N}[(y^{**}-x^{**})\nonumber\\
&+&(1-y^{**})-(1-x^{**})]\nonumber\\
&=&0.
\end{eqnarray*}
\setlength{\arraycolsep}{5pt}

We have $q_{2}^{2}-4p_{2}<0$ and $q_{2}=0$, therefore the eigenvalues of the Jacobian matrix are pure imaginary. The dynamics analysis of the interior of $S_{3}$ and the stability of interior fixed point can be found in the following subsection.

\subsection*{D.3 The Hamiltonian system}\label{AppendixA.42}

To analyze the dynamics in the interior of $S_{3}$, we introduce a new variable $\varepsilon=\frac{x}{x+y}$, which represents the fraction of cooperators among members who do not contribute to the exclusion pool. This yields
\begin{eqnarray}
\dot{\varepsilon}&=&\frac{xy}{(x+y)^{2}}(P_{C}-P_{D})\nonumber\\
&=&-\varepsilon(1-\varepsilon)(P_{D}-P_{C}).
\end{eqnarray}
By substituting $x=\varepsilon(1-z)$ and $\bar{P}=x(P_{C}-P_{D})+(1-z)(P_{D}-P_{F})+P_{F}$ into $\dot{z}=z(P_{F}-\bar{P})$, we have $\dot{z}=z[x(P_{D}-P_{C})-(1-z)(P_{D}-P_{F})]$.
Thus we have
\setlength{\arraycolsep}{0.0em}
\begin{eqnarray}
\left\{
\begin{aligned}
\dot{\varepsilon}&=-\varepsilon(1-\varepsilon)[(1-z)^{N-1}\frac{rc(N-1)}{N}-rc+c],\label{function3}\\
\dot{z}&=z(1-z)[rc-c-\delta-\tau-\varepsilon(rc-c)].
\end{aligned}
\right.
\end{eqnarray}
\setlength{\arraycolsep}{5pt}
By dividing the right-hand side of Eq. (\ref{function3}) by the function $\varepsilon(1-\varepsilon)z(1-z)$, we further have
\setlength{\arraycolsep}{0.0em}
\begin{eqnarray}
\left\{
\begin{aligned}
\dot{\varepsilon}&=-\frac{1}{z(1-z)}[(1-z)^{N-1}\frac{rc(N-1)}{N}-rc+c],\\
\dot{z}&=\frac{1}{\varepsilon(1-\varepsilon)}[rc-c-\delta-\tau-\varepsilon(rc-c)].
\end{aligned}
\right.
\end{eqnarray}
\setlength{\arraycolsep}{5pt}
Let us introduce $H(\varepsilon,z)=M(z)+L(\varepsilon)$, where $M(z)$ and $L(\varepsilon)$ are primitives of $\dot{z}$ and $\dot{\varepsilon}$, which are respectively given as
\setlength{\arraycolsep}{0.0em}
\begin{eqnarray*}
M(z)&=&(-\delta-\tau)\log(1-\varepsilon)-(rc-c-\delta-\tau)\log(\varepsilon),\\
L(\varepsilon)&=&(rc-c)[\log(1-z)-\log(z)]\\
&+&\sum_{k=1}^{N-2}\binom{N-2}{k}(-1)^{k}\frac{z^{k}}{k}+\log(z).
\end{eqnarray*}
\setlength{\arraycolsep}{5.0pt}
Then we obtain the Hamiltonian system as
\begin{eqnarray}
\left\{
\begin{aligned}
\dot{\varepsilon}=\frac{\partial H}{\partial z},\\
\dot{z}=\frac{\partial H}{\partial \varepsilon}.
\end{aligned}
\right.
\end{eqnarray}

Thus the system is conservative, and all constant level sets of $H$ are closed curves.


\section*{Acknowledgments}

This research was supported by the National Natural Science
Foundation of China (Grant No. 61503062) and by the Slovenian Research Agency (Grant Nos. J1-7009 and P5-0027).

\end{document}